\documentclass[a4paper,twocolumn,11pt]{quantumarticle}
\pdfoutput=1
\usepackage{epsfig}
\usepackage{amssymb}
\usepackage{amsmath}
\usepackage{amsfonts}
\usepackage{braket}
\usepackage[T1]{fontenc}
\usepackage{bm}
\usepackage{graphicx}
\usepackage{xcolor}
\usepackage{verbatim}
\usepackage{soul}
\usepackage{float}
\usepackage{tikz}
\usetikzlibrary{matrix,shapes,arrows,positioning,chains}
\usetikzlibrary{shapes.geometric}
\usetikzlibrary{shapes.arrows}
\usepackage{array}
\usepackage{enumitem}
\usepackage{color}
\usepackage[normalem]{ulem}
\usepackage{physics}
\usepackage{caption}
\usepackage{subcaption}
\usepackage[export]{adjustbox}
\usepackage[percent]{overpic}

\usepackage{hyperref}

\begin{document}

\title{Fully optimised variational simulation of a dynamical quantum phase transition on a trapped-ion quantum computer}

\author{Lesley Gover }
\affiliation{London Centre for Nanotechnology, University College London, Gordon St., London, WC1H 0AH, United Kingdom}

\author{Vinul Wimalaweera}
\affiliation{London Centre for Nanotechnology, University College London, Gordon St., London, WC1H 0AH, United Kingdom}

\author{Fariha Azad}
\affiliation{London Centre for Nanotechnology, University College London, Gordon St., London, WC1H 0AH, United Kingdom}

\author{Matthew DeCross}
\affiliation{Quantinuum, Broomfield, CO 80021, USA}

\author{Michael Foss-Feig}
\affiliation{Quantinuum, Broomfield, CO 80021, USA}

\author{A.~G. Green}
\affiliation{London Centre for Nanotechnology, University College London, Gordon St., London, WC1H 0AH, United Kingdom}
\affiliation{email: andrew.green@ucl.ac.uk}

\begin{abstract}
We time-evolve a translationally invariant quantum state on the Quantinuum H1-1 trapped-ion quantum processor, studying the dynamical quantum phase transition of the transverse field Ising model. This physics requires a delicate cancellation of phases in the many-body wavefunction and presents a tough challenge for current quantum devices. We follow the dynamics using a quantum circuit matrix product state ansatz, optimised for the time-evolution using a fidelity cost function. Sampling costs are mitigated by using the measured values of this circuit as stochastic corrections to a simple classical extrapolation of the ansatz parameters.
Our results 
demonstrate the feasibility of variational quantum time-evolution and 
reveal a hitherto hidden simplicity of the evolution of the transverse-field Ising model through the dynamical quantum phase transition.
\end{abstract}
\maketitle
\tableofcontents

\vspace{-0.2in}

 \section{Introduction}
 \label{sec:introduction}
Many prominent advances in modern condensed matter physics rely on variational
ansatze. They were key to findings such as the BCS theory of superconductivity, the quantum Hall effect and spin
liquids \cite{bardeenTheorySuperconductivity1957, klitzingNewMethodHighAccuracy1980, savaryQuantumSpinLiquids2016}. Moreover, some foundational numerical methods in the study of strongly correlated systems also use variational ansatze. The density matrix renormalization group (DMRG) algorithm utilises matrix product states (MPS) revealing the importance of entanglement structure in the many-body wavefunction \cite{SteveWhiteDMRG, OstlundRommer, mps_representations, schollwock_review}.

Variational quantum algorithms are a leading strategy to simulate quantum systems with noisy intermediate scale (NISQ) devices.
Core to
these algorithms is the choice of the variational ansatz
\cite{cerezoVariationalQuantumAlgorithms2021}. The structure of the variational ansatz will
determine the rate at which the algorithm converges, the accuracy of the solution and the
feasibility of running the algorithm.
Therefore, careful consideration has to be made of the choice of ansatz.

Tensor networks are a good choice for a variety of problems. Classically, they are
state of the art for a number of interesting problems in quantum simulation and in fields ranging from fluid dynamics to machine learning. Furthermore, there is a natural translation from tensor networks to quantum circuits \cite{schon2005sequential, circuits_are_mps, Smith2019, Lin2021real, barratt2021parallel}, making them good 
ansatze for use in quantum algorithms. 
Quantum tensor networks can have an exponential advantage in run time compared to the time required to contract the classical tensor network\cite{Lin2021real}. The classical contraction time scales polynomially with the bond dimension, whereas the individual unitaries of the quantum representation can have a depth that is logarithmic in the bond dimension, leading to a logarithmic run time.

Even with a good ansatz, sampling costs can be a fundamental limitation of variational algorithms; they can quickly overwhelm any quantum advantage \cite{wecker2015progress}. 
Measurements can be the slowest process in a computation, and the sampling needed to attain the required accuracy quickly becomes intractable. 
Utilising variational algorithms for applications such as time evolution, where an optimisation loop is often required at each time step, seems
infeasible. However, good initialisation of the variational optimisation can give an exponential reduction in sampling. 

Here, we perform variational updates, harnessing features of time-evolution, to dramatically reduce sampling costs. We implement this using Quantinuum's H1-1 device to perform time-evolution with a translationally invariant matrix product state (iMPS) ansatz.
Time-evolution is captured by smooth changes of the variational parameters \cite{puig2024variationalquantumsimulationcase}. Classical extrapolation — for example, with polynomial splines — corrected stochastically using calls to the quantum computer 
increases sampling efficiency by several orders of magnitude.

\section{Results}
\begin{figure*}
  \begin{center}
    \begin{subfigure}{0.75\textwidth}
        \caption{}
        \includegraphics[width=\textwidth]{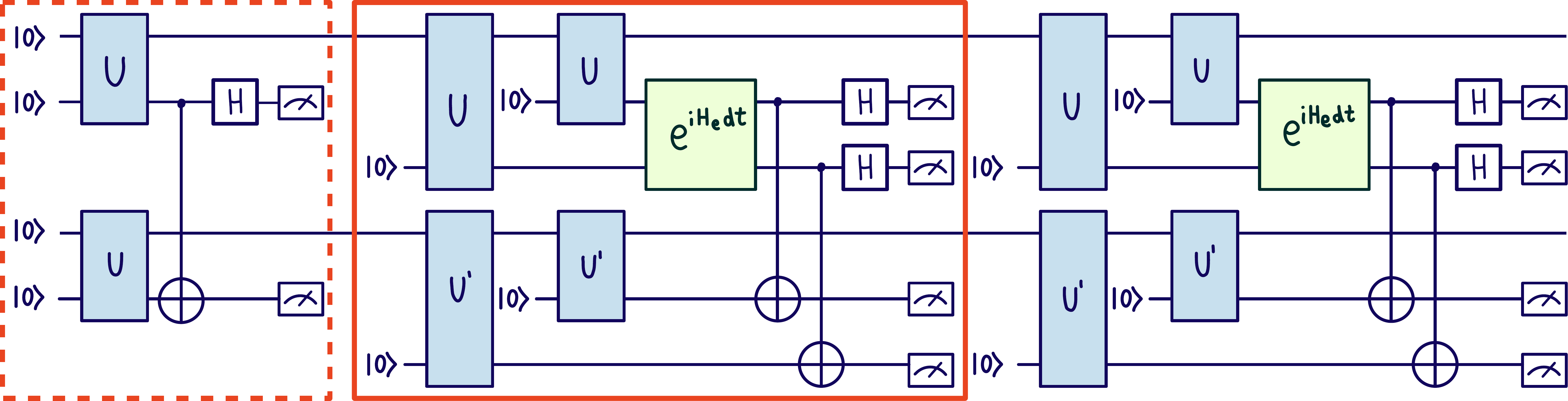}
    \end{subfigure}
    \begin{subfigure}{0.5\linewidth}
        \caption{}
        \includegraphics[width=\linewidth]{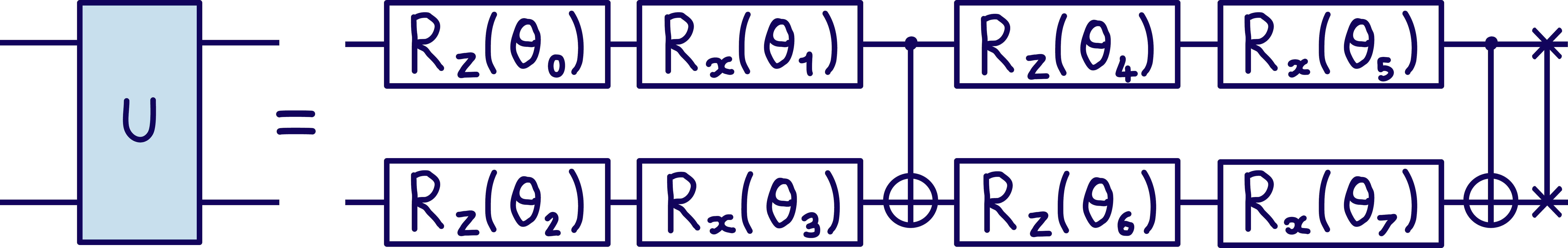}
    \end{subfigure}
    \begin{subfigure}{.85\linewidth}
        \caption{}
        \includegraphics[width=\linewidth]{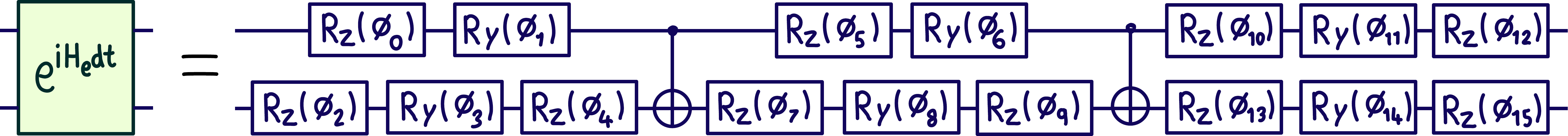}
    \end{subfigure}
  \end{center}
  \caption{ {\bf a) Cost Function Circuit.} Optimisation of this circuit over $U'$ gives the unitary describing the time evolution of the state unitary $U(t)$ to the next time step. This circuit is a second order power method evaluation of the leading eigenvalue of the mixed transfer matrix (indicated by the solid red box). The right fixed point of this transfer matrix is approximated to $O(dt^2)$ by the open circuit. The left fixed point (indicated by the dashed red box) is approximated as shown - the optimum fidelity density is not affected by this choice. 
  {\bf b) Ansatz decomposition.} The two qubit unitary $U$ representing the matrix product state is approximated by the reduced ansatz shown. This reduces the number of parameters to be optimised for our experiments at the cost of reducing its theoretical optimum performance slightly. 
  {\bf c) The time-evolution Unitary } is compiled to the gate set as shown, where $\{\phi_0,\phi_1,\phi_2,\phi_3,\phi_4,\phi_5,\phi_6,\phi_7,\phi_8,\phi_9,\phi_{10},\phi_{11},\phi_{12},\phi_{13},\phi_{14},\phi_{15}\}=\{-0.5\pi, 0.987\pi, \pi, 0.5\pi, 0.513\pi, 0.5\pi,-\pi, -0.5\pi, 0.5, \pi, 0.245\pi, -0.5\pi, 0.013\pi, -0.99\pi, 0.013\pi\}$.}
\label{fig:circuit}
\end{figure*}

We report simulations of the dynamical quantum phase transition in the transverse-field Ising model \cite{sachdev2011quantum} on Quantinuum's H1-1 quantum computer \cite{H11}.
 We begin with a brief discussion of this target problem. Next, we introduce the translationally invariant quantum circuit infinite matrix product state (iMPS) ansatz \cite{barratt2021parallel, dborinSimulatingGroundstateDynamical2022} optimised to make use of the mid-circuit measurement facility of the H1-1 device, and our algorithm to perform its variational time-evolution.
 Finally, we present results demonstrating that variational evolution of this ansatz on Quantinuum's H1-1 device captures the key features of the dynamical quantum phase transition.

\vspace{0.1in}
{\bf Dynamical Quantum Phase Transition:}
The transverse-field Ising model is a well-understood model with some subtle features in its statics and dynamics that provide a challenging test for current quantum devices  \cite{sachdev2000quantum, sachdev2011quantum}. Its Hamiltonian is given by

\begin{equation}\label{eq:TFIM}
    H = \sum _i [J\hat{Z}_i\hat{Z}_{i+1} + g\hat{X}_i],
\end{equation}
where $\hat{Z}$ and $\hat{X}$ are the Pauli operators, $J$ the exchange coupling, and $g$ the magnitude of the transverse field. This model harbours a quantum phase transition in its ground state at $g/J = 1$. A dynamical quantum phase transition is seen when a ground state is prepared on one side of this critical point and then evolved with a Hamiltonian with $g/J$ on the other side of it \cite{pollmann2010dynamics, dynamical_phase_transitions, dpt_observing, dpt_probing}. This is seen as recurrences and cusps in the Loschmidt echo — the negative log of the overlap between the initial state and the evolved state: $-\log|\langle\psi(0)|\psi(t)\rangle|$. The dynamical quantum phase transition corresponds to the cusps in the Loschmidt echo. Seeing them in experiment involves a subtle cancellation of phases in the many-body wavefunction.

\vspace{0.1in}
{\bf Ansatz and Time-Evolution:}
The basis of our algorithm is to perform time evolution by iteratively updating a translationally invariant quantum circuit iMPS \cite{dborinSimulatingGroundstateDynamical2022}. At each time-step, the ansatz is updated using a Trotterised evolution and projected back onto the variational manifold by optimising a fidelity cost function:
\begin{equation}
    U(t+dt) = \arg \max_W |\bra{\psi(W)} e^{-iHdt} \ket{\psi(U(t))}|^2,
    \label{eq:FidelityCost}
\end{equation}
where the state at time $t$ is described by a sequence of two-qubit unitaries $U(t)$, and $U(t+dt)$ gives the update to this state at time $t+dt$. 
Strictly, the fidelity is zero for non-identical translationally invariant states. Because of this, in practice the fidelity density is optimised. 
A number of choices and approximations must be balanced in the evaluation of this: the bond-dimension of the iMPS; the timestep (this must be large enough to give a sufficient resolution but not too large as to incur unacceptable Trotter errors); the order of Trotterisation; the order of power method used to determine the fidelity density [see Methods for more details]. The optimal choices depend upon capabilities of the particular quantum device, offsetting errors introduced by a particular approximation against reduced exposure to device infidelities and sampling costs. 

The quantum circuit that we use to approximate the fidelity density is shown 
in Fig.~\ref{fig:circuit}. 
Each unitary is compiled to the reduced gate set given in Fig.~\ref{fig:circuit}b) in order to reduce the number of parameters to be optimised at each timestep. 
To further reduce our sampling costs, we used a linear extrapolation of the variational parameters to 
make a good initial estimate of the next time step. Optimising the cost function on the quantum computer can be seen as making stochastic corrections to this parameter update. 
Further details of the cost function and update algorithm  are given in \ref{sec:Methods}.

\begin{figure}[t]
\includegraphics[width=0.5\textwidth]{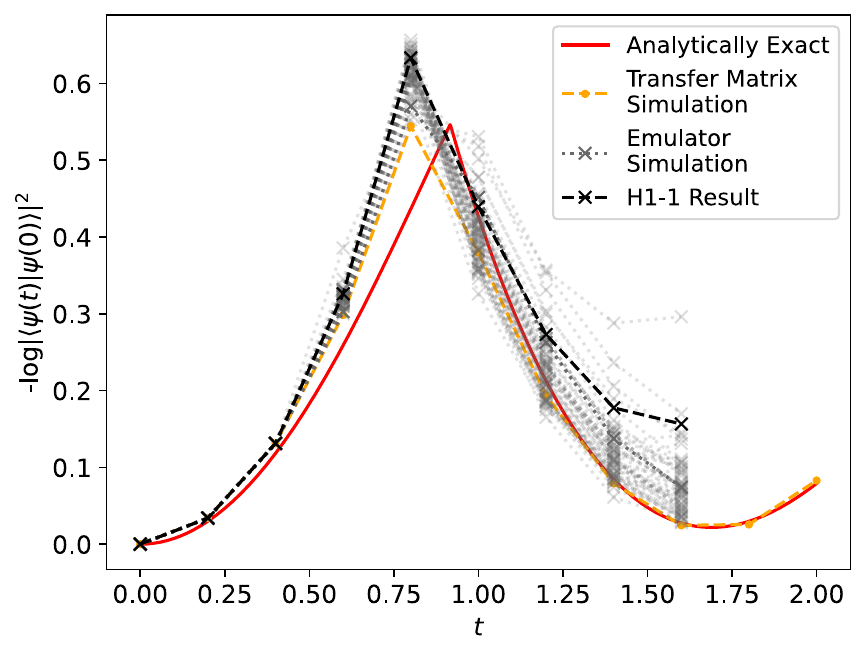}
\caption{{\bf Simulating a Dynamical Quantum Phase Transition on Quantinuum H1-1}.
  These data show the Loschmidt echo, $- \log |\langle \psi(t)| \psi(0) \rangle |^2$, for a quench of the transverse-field Ising model Eq.(\ref{eq:TFIM}) across its quantum critical point. The ground state for $g=1.5$ is evolved with $g=0.2$.
This evolution displays recurrences where the evolved state becomes close to the initial state and dynamical quantum phase transitions in between. The exact evolution is known and shown by the dashed red curve. Using a quantum circuit MPS ansatz, with a bond dimension $D=2$ an exact time-dependent variational principle evolution is shown by the orange dashed curve. The grey curves show 100 emulated runs and the solid black curve the result of a run on the Quantinuum H1-1 device. The spread of these results is largely determined by sampling errors.}
\label{fig:CentralResults}
\end{figure}

\vspace{0.1in}
{\bf Experimental Results:}
The results of implementing our algorithm
on Quantinuum's H1-1 device  are shown in Figure \ref{fig:CentralResults}. 
These data show the Loschmidt echo for the dynamical phase transition starting with $g=1.5$ and evolving with $g=0.2$. The optimisation of the cost function is performed using the SPSA optimiser \cite{spall1992multivariate}. The results show a single full run on the quantum computer overlaid on 50 runs on the H1-1 emulator. The Loschmidt echo shows up well and has a good agreement with the results from the H1-1 emulator. The variance in the emulator runs is a result of machine infidelity and sampling error, and is dominated by the sampling error, as shown in Appendix \ref{app:SamplingErrors}. 

A surprising feature of our results is that the variational parameters change rather linearly with time, suggesting that the dynamical quantum phase transition can be understood as a precession in the space of iMPS parameters. This seems to be a lucky accident of our choice of our ansatz,
which admits both 
 reparametrisations that describe (approximately but with high fidelity) the same state in the same gauge, and reparametrisations that describe a restricted set of exact gauge transformations of the state. 

\section{Methods} \label{sec:Methods}
{\bf iMPS Ansatz and fidelity density cost function}
The iMPS fidelity in principle requires contracting a spatially infinite state and consequently an infinite circuit.
However, as shown in Refs.~\cite{barratt2021parallel, dborinSimulatingGroundstateDynamical2022}, it is possible to evaluate this fidelity with a finite circuit. The
fidelity cost function given by Eq.~\ref{eq:FidelityCost} corresponds to an infinite circuit
shown in Fig.\ref{fig:circuit}. In
this figure, the evolution operator is implemented using a first order Trotterisation with a trick noted in Ref.~\cite{dborinSimulatingGroundstateDynamical2022} applicable to translationally invariant states. The iMPS
ansatz circuit has a bond order of 2. Formally, the fidelity between the evolved state $e^{-iHdt}
\ket{\psi(t)}$ and the variational ansatz state $\ket{\psi(W)}$ is zero when they are not identical, since the states are translationally invariant.
The fidelity density $\lambda$ captures the rate at which the fidelity between the two states
approaches zero and is given by the leading order of the mixed transfer matrix highlighted in solid red in Fig.~\ref{fig:circuit}. Therefore, to optimise the fidelity cost function, an efficient
way of calculating $\lambda$ is necessary.

The fidelity density $\lambda$ can be found using the power method. The power method states that the
leading order eigenvalue of the transfer matrix $E_{A, B}$ is given by
\begin{equation}
  \lambda = \lim_{n \rightarrow \infty} \frac{\widetilde{L} (E_{A, B})^n
  \widetilde{R}}{\widetilde{L} (E_{A, B})^{n-1} \widetilde{R}} = \frac{C_n(A, B)}{C_{n-1}(A, B)},
  \label{eq:power_method}
\end{equation}
where $\widetilde L$ and $\widetilde R$ are approximations to the left and right eigenvectors of the transfer matrix.
Depending upon the quality of $\widetilde L$ and $\widetilde R$ the power method actually converges with fairly small $n$.
When either $\widetilde L$ and $\widetilde R$ are exactly the left and right eigenvectors of $E_{A, B}$ the power method works at
$n=1$.

Implementing the power method requires embedding the
approximation of $\widetilde L$ and $\widetilde R$ onto gates in the circuit. This can be done numerically by solving
a fixed point equation through variational optimisation, however this is prohibitively expensive as
it requires an internal optimisation loop. As we are time evolving the state $\ket{\psi(U(t))}$ with
small time steps, 
$\widetilde R$ is approximated to order $dt^2$ by the right fixed point of the transfer matrix $E_{U,U}$. This is given by the identity and realised in a Hadamard test by the open circuit configuration on the auxiliary qubits as shown in Fig.(\ref{fig:circuit}) [see also Appendix.\ref{app:CostFunction}].
We use a simple approximation for $\widetilde L$ using two copies of the current state unitary $U(t)$.

Our results obtained on Quantinuum's H1-1 device used the second order
of the power method. Fig.~\ref{fig:circuit} gives the circuit for implementing
$C_2(A, B)$ with these approximate $\widetilde L$ and $\widetilde R$. These circuits are a remapping of the staircase
`space-like' cost function outlined in Ref.~\cite{dborinSimulatingGroundstateDynamical2022}. This `space-like' cost
function was well suited to the superconducting Sycamore device, which has access to a higher number of
qubits but with short coherence times. This `time-like' rewrite utilises the longer coherence times and access to mid-circuit measurement and reuse
of qubits available on Quantinuum's H1-1 device. These time-like circuits are referred to in literature variously as the holographic representation of an MPS \cite{holographicQuantinuum2021, gopalakrishnanUnitaryCircuitsFinite2019, kimNoiseresilientPreparationQuantum2017} or sequential circuits \cite{PhysRevLett.95.110503,PhysRevA.75.032311}.

In addition to laying out the circuit in a time-like fashion, further choices are made based upon
device performance. 
The data that we collected on H1-1 used a first order Trotterisation. However, the high gate fidelity and the fact that the accuracy of our results was sampling limited suggests that a higher order Trotterisation and larger $dt$ may be favourable [see Appendix].
Note that the circuit shown in Fig.\ref{fig:circuit} uses a trick developed in Ref.\cite{dborinSimulatingGroundstateDynamical2022} whereby combining first order Trotterisation and projection back onto the iMPS
circuit manifold only requires implementation of the even part of the Trotterisation (scaled by a factor of 2). This is evident from the time-dependent variational principle, whose equations are invariant under this change. 

{\bf Stochastic Correction of Classical Extrapolation:}
Our time evolution algorithm requires a variational optimisation at each step in the evolution. Generally, such optimisation can be prohibitively expensive, offsetting advantage in contracting the circuit on a quantum device. The cost of this optimisation
can be significantly reduced given a good estimate for the initial parameters. Fortunately, in time
evolution, we expect variational parameters to change continuously. Therefore, given information about the values of the parameters at previous time steps, a good initial guess
can be made for the optimisation of the cost function using a simple extrapolation \cite{puig2024variationalquantumsimulationcase}.

The extrapolation of variational parameters can be performed efficiently classically. In
principle, sophisticated methods such as high order polynomial splines or machine learning models can be used for complicated parameter evolutions. 
Here, we use a simple linear extrapolation 
from the two previous time steps to seed the optimisation results with more than three orders of
magnitude reduction in the number of shots required [see Appendix for further details]. This turns our algorithm from intractable to feasible on near-term quantum devices. Viewed in this way, the quantum device provides stochastic refinement of a classical
probabilistic model. 

\section{Discussion}
 We have demonstrated a full
variational time evolution algorithm using
a translationally invariant tensor network ansatz on a NISQ device. Using this algorithm, we were
able to capture the dynamical quantum phase transition 
 of the transverse field Ising model. This subtle, many-body quantum effect is a challenging 
  test of the current generation of quantum devices. In doing so, we found
that Quantinuum's H1-1 device showed excellent fidelity, requiring no error mitigation for
depolarising noise. The main restriction to our algorithm is sampling complexity.

A key contribution made here is to mitigate these costs using an optimised sampling routine. 
Variational parameters change smoothly during quantum time evolution, enabling us to seed the optimisation at each time step using simple extrapolation from previous time
steps. This resulted in several orders of magnitude reduction in sampling costs, rendering this algorithm feasible given current device performance. It is an improvement over current warm-start techniques which copy parameters from previous time steps to avoid issues in training such as barren plateaus \cite{puig2024variationalquantumsimulationcase}. To achieve quantum
advantage with MPS algorithms on quantum devices relies on the exponential reduction in
qubits needed to represent an arbitrary bond dimension $D$ \cite{Lin2021real}. Despite a simple linear extrapolation
working well for a bond dimension $D=2$, more sophisticated parameterisation and extrapolation
methods may be required as this is scaled.  More sophisticated machine learning models have seen success in other areas of quantum simulation and control and may be useful here \cite{rodriguez2019identifying, carrasquilla2017machine, metz2022self}.

Current NISQ devices have a broad range of architectures and characteristics to which near-term algorithms must adapt \cite{bhartiNoisyIntermediatescaleQuantum2022}. Our approach has several control parameters that can be tuned. For example, circuit structure (space-like or time-like) can be chosen to accommodate coherence times and whether mid-circuit measurements are supported. The size of the time steps and Trotterisation lead to tradeoffs where larger time steps are easier to resolve but require higher order Trotterisation with greater exposure to gate errors. Choosing a higher order in the power method permits more accurate determination of the fidelity density, but again increases sensitivity to gate errors.  
Each device will have a unique sweet spot or signature of tunable parameters locating the optimum tradeoff.
Hence, the algorithm is readily portable, subtly adapting to particular device characteristics \footnote{Indeed, this portability is demonstrated by the fact that the cost function for time evolution was first
profiled ({\it mutatis mutandis}), but not optimised, on Google's Sycamore superconducting architecture \cite{dborinSimulatingGroundstateDynamical2022}.}

This work admits several directions of exploration in the pursuit of quantum advantage.
Firstly, the advantageous scaling for quantum MPS algorithms is as a function of the bond dimension. 
Developing techniques for increasing bond dimension whilst maintaining the feasibility of
running this algorithm of near-term devices is necessary. Scaling with bond dimension is more severe  --- and the potential for quantum advantage greater --- in higher dimensions. Adapting the present approach to 2D and higher dimension models is desirable. This could be achieved either by performing a raster of an MPS over the higher dimension or utilising sequential circuits directly
to generate subclasses of PEPs \cite{wei2022sequential, banuls2008sequentially, isometric_peps}. 
Finally, directly evolving classical MPS using algorithms such as TDVP is inherently
limited due to ballistic growth in entanglement and the corresponding exponential growth in bond dimension required to encode the wavefunction.
However, motivated by the notion that local properties or correlators of local properties should scale with intensive quantities, recent tensor network algorithms \cite{PhysRevB.97.035127,hallam2019lyapunov,PhysRevX.8.021013,PhysRevB.105.075131} have demonstrated that long-time evolution of local observables is possible in thermalising quantum systems. It would be revealing to use the methods developed here to translate these algorithms to run on quantum computers. 

 Simulations such as the one achieved here would have been infeasible a few years ago.
The competition between quantum algorithms and classical tensor network simulations has been a potent driver of this scientific advance 
\cite{48651,PRXQuantum.4.020304,Gray2021hyperoptimized,PhysRevLett.129.090502,IBMUtility,PRXQuantum.5.010308,doi:10.1126/sciadv.adk4321}.
Translating tensor network code to run on quantum computers has the potential to accelerate this interplay. The structure of the quantum tensor network circuits directly mirror classical
tensor network models.
Quantum advantage arises through quantum run times that scale logarithmically with bond dimension compared to classical contraction times that scale polynomially with bond dimension.
Moreover, tensor networks are not limited to quantum simulation \cite{doi:10.1137/090752286,Gourianov:2021tot}. Quantum computing is on the cusp of quantum advantage \cite{decross2024computationalpowerrandomquantum,andersen2024thermalizationcriticalityanalogdigitalquantum} and tensor networks have an important role to play in delivering on this promise.

\section{Acknowledgements} We acknowledge funding from the EPSRC through EP/S021582/1.

\bibliographystyle{naturemag}
\bibliography{Bibliography.bib}

\appendix
\section{Cost Function}
\label{app:CostFunction}
The cost function circuit in Fig.\ref{fig:circuit} can be obtained as a representation of the fidelity cost function Eq.(\ref{eq:FidelityCost}) as follows:

\noindent
i. Without qubit reuse, the fidelity between the time-evolved state and an updated MPS ansatz can be represented as an infinite circuit
\begin{eqnarray*}
&&\langle \psi(U') |e^{-i H dt}  | \psi(U) \rangle\\
&=&
\raisebox{-0.6in}{
\includegraphics[width=0.8\linewidth]
{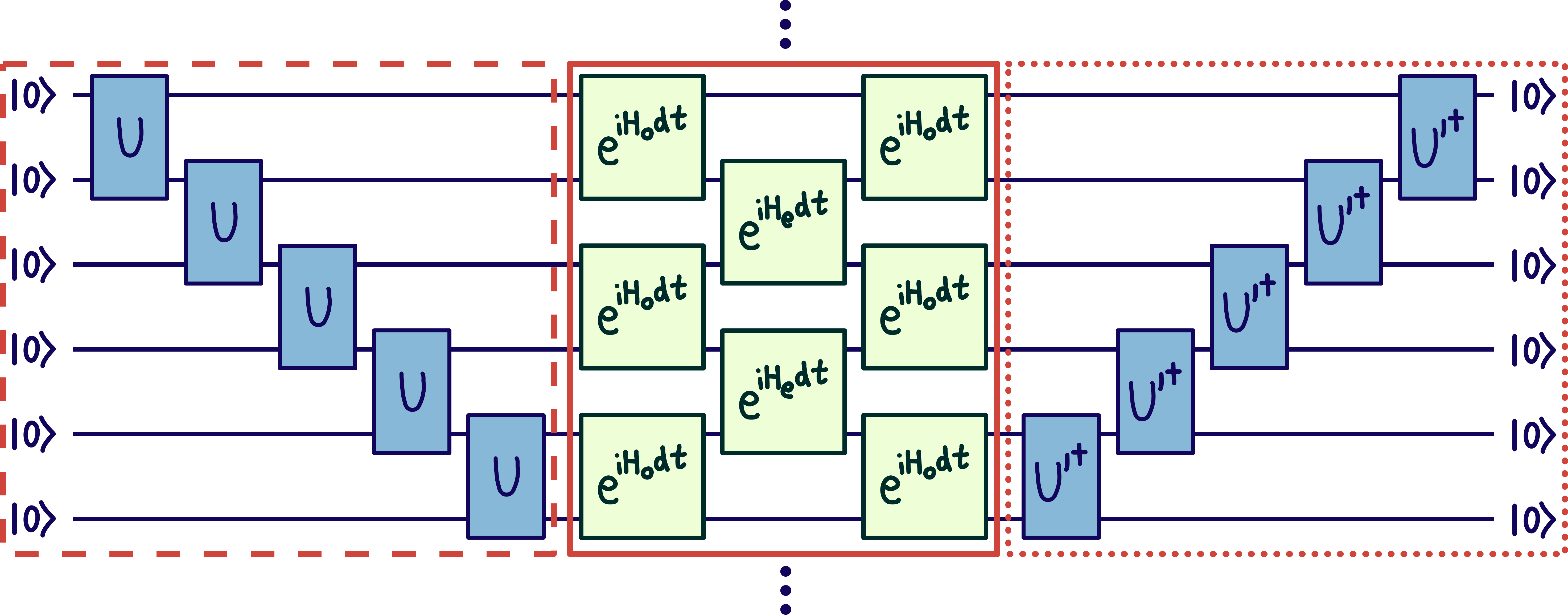}}.
\end{eqnarray*}
This can be simplified by realising that for translationally invariant states we get the same update using just the even terms in the Hamiltonian (with appropriate rescaling of the time-step);
\begin{eqnarray*}
&&\langle \psi(U') |e^{-i H dt}  | \psi(U) \rangle\\
&\approx&
\raisebox{-0.6in}{
\includegraphics[width=0.8\linewidth]
{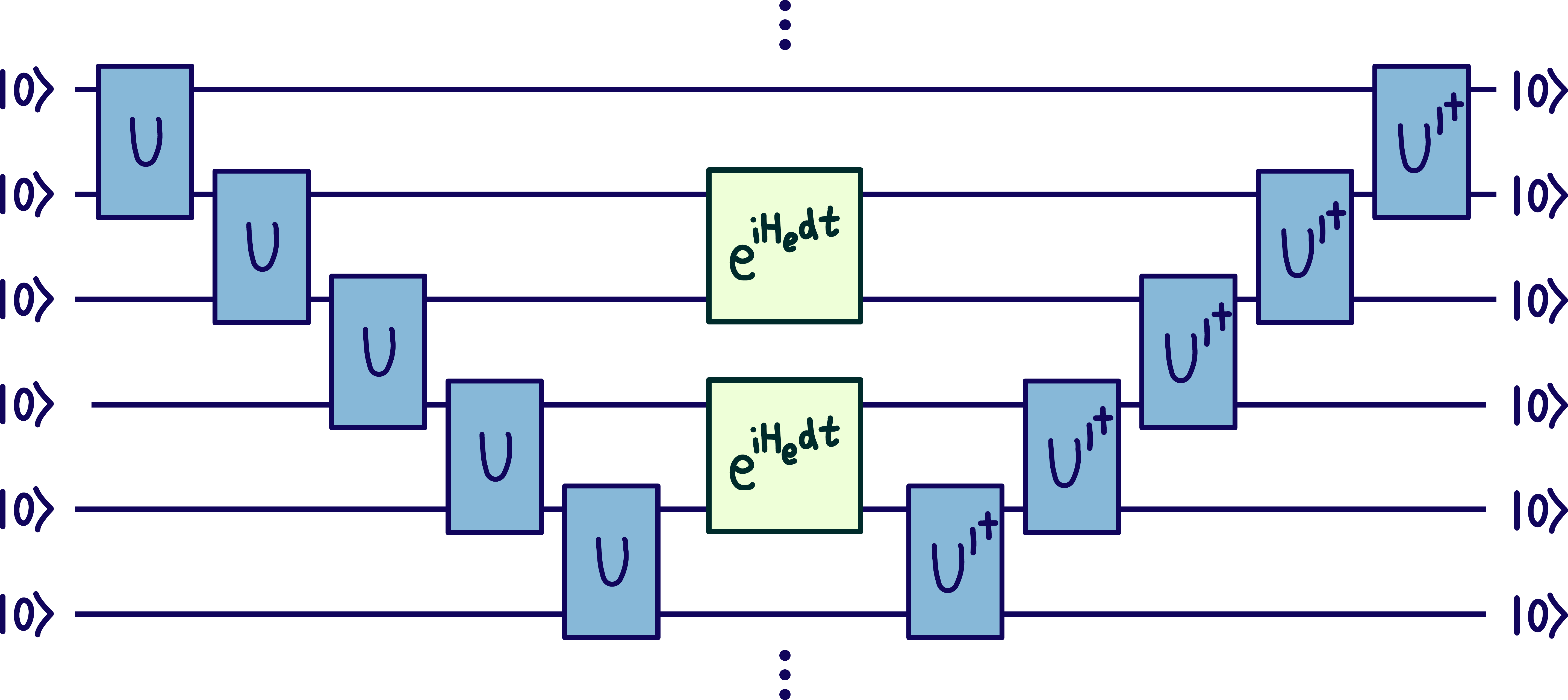}}
,
\end{eqnarray*}

ii. This is strictly zero in the thermodynamic limit and, as discussed in the main text, we calculate a second order power method approximation to the leading eigenvalue of the transfer matrix. Moreover, we choose simple approximations for the fixed points of this transfer matrix; 
$$
\lambda^2
\approx
\raisebox{-0.6in}{
\includegraphics[width=0.8\linewidth]
{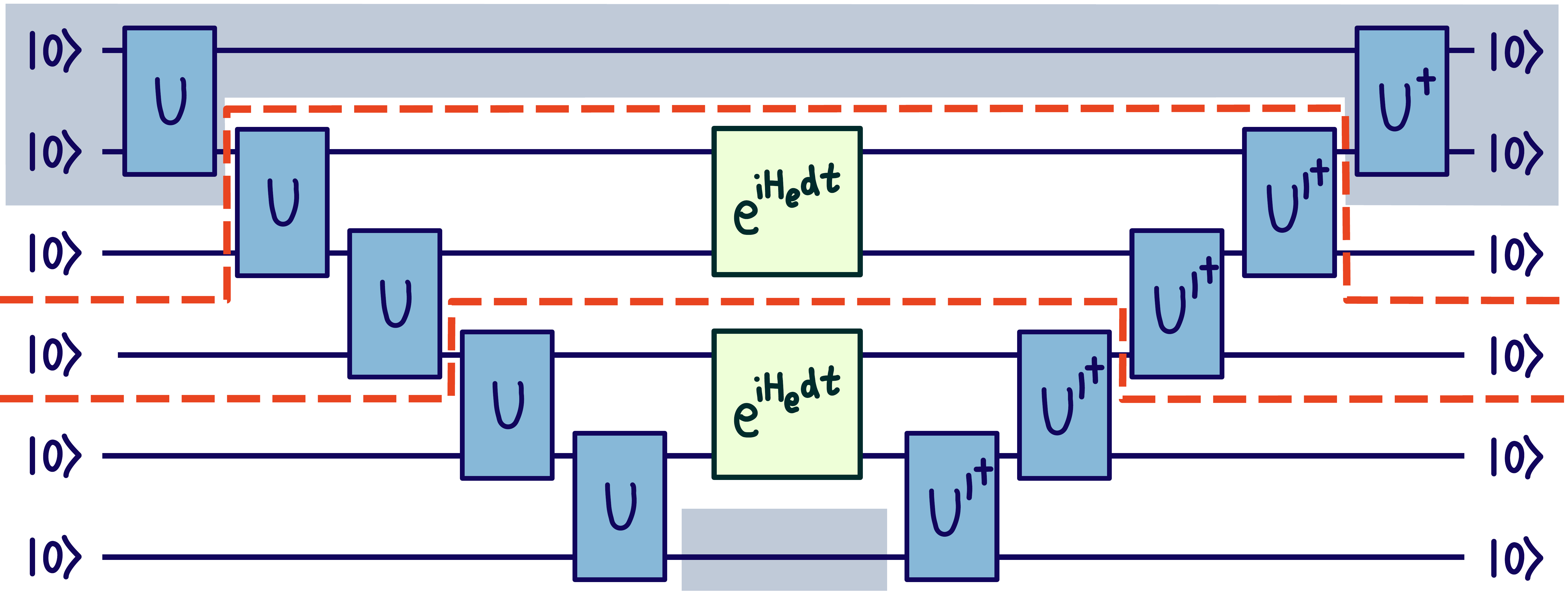}}
,
$$
where the red dashed lines indicate the transfer matrix and the grey shading indicates the approximate fixed points of this transfer matrix. The latter are the fixed points of the transfer matrix used to evaluate the expectation in a translationally invariant state described by the unitary $U$. 

iii. We can equivalently write this circuit using a Hadamard test \cite{swap_test};
\begin{widetext}
\begin{eqnarray*}
|\lambda|^4
\approx
\raisebox{-1.1in}{
\includegraphics[width=0.4\linewidth]
{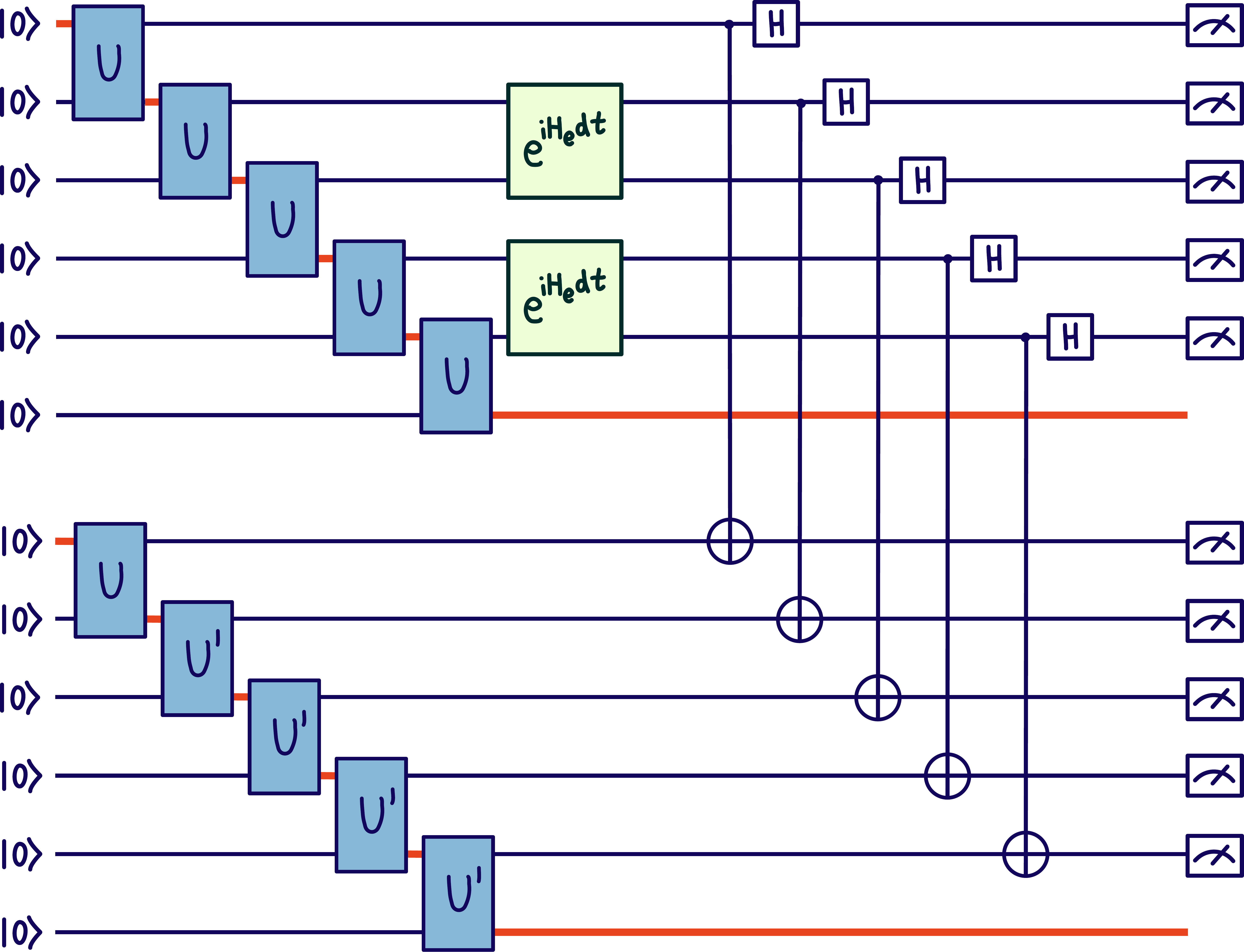} },
\end{eqnarray*}
where we have highlighted the auxiliary space qubit lines in red. 
The lack of Hadamard test on the auxiliary space qubits is a property of the infinite sequential nature of our states. We shall turn to this in a moment.

iv.
Recasting the above circuit by reusing qubits recovers Fig.(\ref{fig:circuit}), where the unitaries in this latter circuit have their outgoing legs swapped.

We are now in a position to address the question of why there is no Hadamard test on the auxiliary space qubits. To do so, we note that approximating the right fixed point (or lower fixed point in step ii.) by the identity is equivalent to extending the circuit with $U$ and $U^\dagger$. In the time-like equivalent circuit, this looks like

\begin{eqnarray*}
|\lambda|^4
\approx
\raisebox{-0.5in}{
\includegraphics[width=0.9\linewidth]
{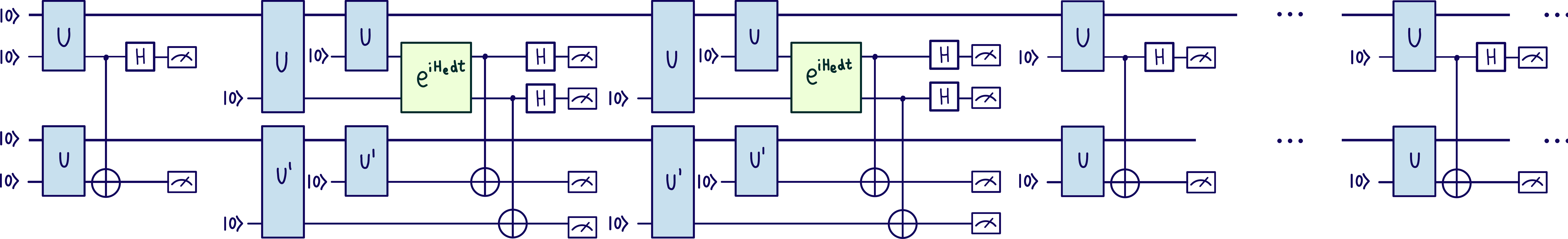}}.
\end{eqnarray*}
The crucial point is that all of the subsequent unitaries are identical in the two states - we get the same result for the overlap if we replace them with the identity;
\begin{eqnarray*}
|\lambda|^4
\approx
\raisebox{-0.5in}{
\includegraphics[width=0.9\linewidth]
{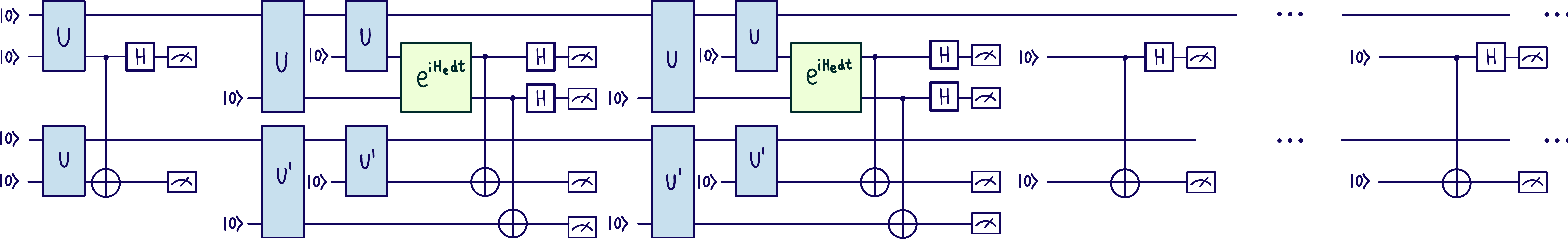}}.
\end{eqnarray*}
From this replacement, we see that measurements on qubits with identical unitaries in the sequential circuit will pass the Hadamard test and so can be excluded from it. No measurement is taken on the auxiliary qubits.  
\end{widetext}

\section{Sampling Errors}
\label{app:SamplingErrors}
Variational ansatze schemes exhibit a significant sampling cost overhead, this can be exponentially high in the worst-case scenario. This can be prohibitive to running these schemes, in particular on trapped-ion devices where the clock times tend to be slower. A method to mitigate this cost is therefore especially important for such devices. Recent work \cite{puig2024variationalquantumsimulationcase} has shown that for time evolutions, a better scaling can be found using the current time step as an initial guess for the optimisation. We have employed a scheme which can do significantly better than this. Our main limitation in running the algorithm was in sampling rather than gate fidelity. A higher order Trotterisation, therefore, may be able to make use of the high fidelity. This is discussed further in Appendix \ref{app:HigherTrotterisation}.

\begin{figure}
    \centering
    \includegraphics[width=0.5\textwidth]{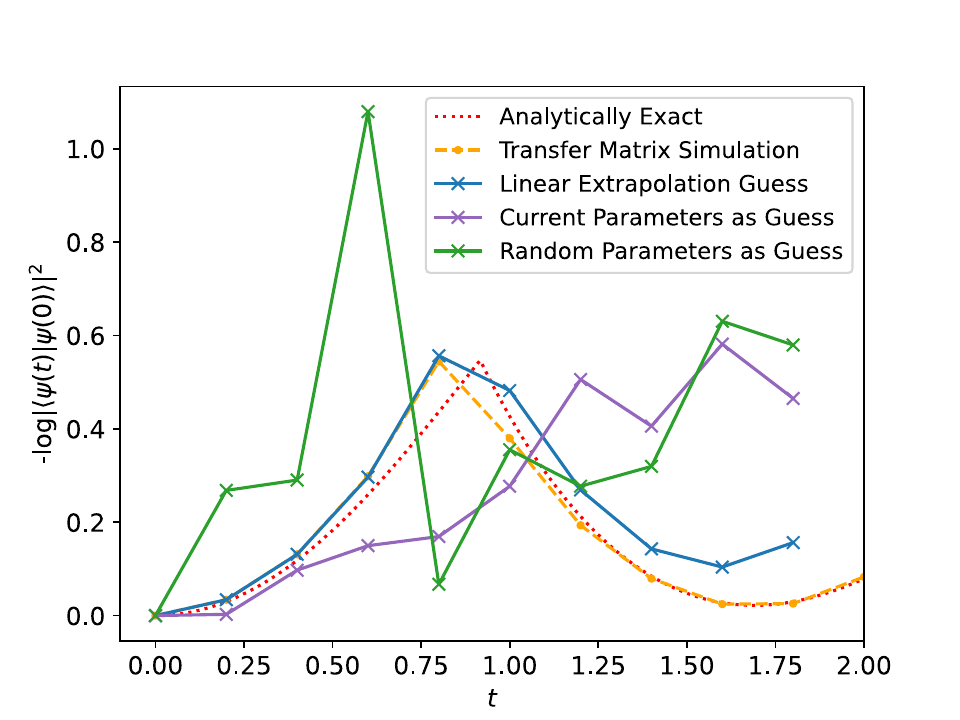}
    \caption{{\bf Performance of time-evolution with different initialisation schemes:} We show the Loschmidt echo calculated using different initialisations of the stochastic optimisation of the parameters for the next time-step. For 6 SPSA steps we use a random initialisation (green curve), initialisation with the parameters of the current time-step (purple curve), and initialisation with linear extrapolation from the previous two time-steps (blue curve). 
    These are compared to analytically exact and exact-in-ansatz results (red and orange curves). The linear extrapolation approximates the exact results with a very small number of SPSA steps. 
    }
\label{fig:ExtrapolationSchemes}
\end{figure}

With no sampling cost mitigation, the full evolution produced would be too costly to run and therefore infeasible. In order to reduce this cost, we use a linear extrapolation in the ansatz parameter space to make an initial guess of the parameters at the next time step. 
This affords a $10^3$ reduction in the sampling space compared to using the current parameters as a starting point for the next time step. Fig. \ref{fig:ExtrapolationSchemes} shows a comparison of how the algorithm performs with the same small number of updates of the SPSA optimisation for 3 different initialisation schemes. We fix the number of SPSA updates to be the same as used in the experiment, and run with perfect gate simulation, \textit{i.e.} so that the only source of noise is sampling error. The figure shows 5 schemes in total: an analytically exact calculation of the Loschmidt echo, a transfer matrix simulation of the ansatz, and perfect gate simulation with random parameters, the current parameters and parameters from a linear extrapolation as an initial guess. From this, it is clear that the linear extrapolation does an excellent job of capturing the dynamics, even with the small update count. It is the only one of the initialisation schemes to reproduce the Loschmidt echo. 

\begin{figure}
    \centering    
    \includegraphics[width=0.45\textwidth]{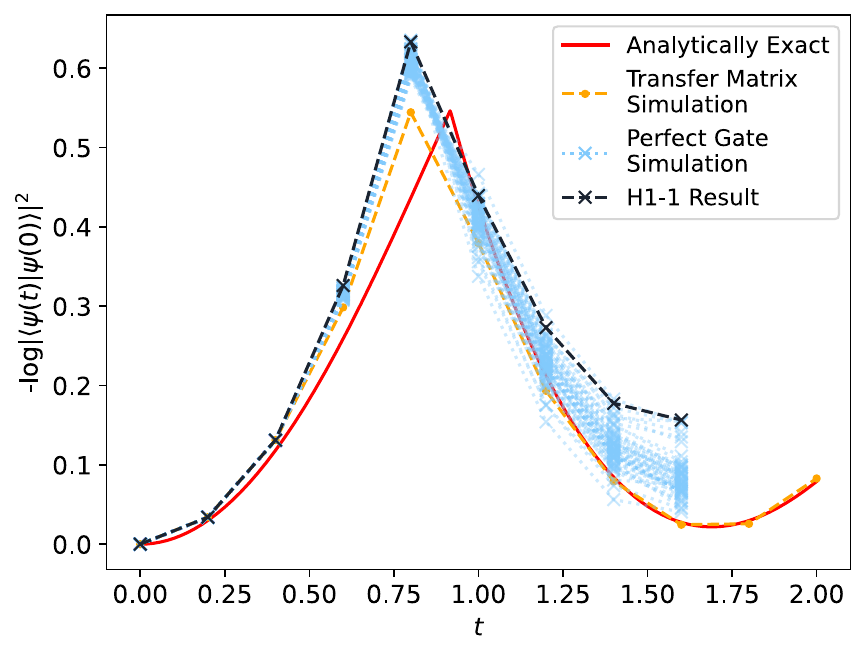}
    \caption{{\bf Performance of time evolution with perfect gate simulation}: The Loschmidt echo produced by simulation on the H1-1 machine (black) compared to 50 runs with perfect gate simulation (blue). The spread in the perfect gate simulation is comparable to that in Fig. \ref{fig:CentralResults} demonstrating that the result is sampling limited.} 
    \label{fig:NoiselessVariance}
\end{figure}

Even with this reduction of the sampling costs, they were still the limiting factor in the performance of the algorithm. Fig. \ref{fig:CentralResults} shows variation towards the end of the set of emulator data. This is due to error introduced as the evolution progresses, and can originate in either the fidelity on the quantum hardware or sampling noise. We find that the main source here is sampling noise. This is demonstrated by simulating with perfect gates but still allowing for sampling error as in Fig. \ref{fig:NoiselessVariance}. It shows that even without gate fidelity error, the variance at the end is similar to that of the emulator data in Fig. \ref{fig:CentralResults}. We therefore conclude that the cause of this variance is sampling error dominated. 

\section{Order of Trotterisation}
\label{app:HigherTrotterisation}
Appendix \ref{app:HigherTrotterisation} suggests that this experiment is significantly sampling limited, device fidelity was not a significant source of error. Given that the optimal operation of this algorithm occurs when various sources of errors are of similar scale, there may be some parameters in this algorithm that can be tuned to get further performance. 

\begin{figure*}
  \begin{center}
  \hspace*{-15cm}(a) \\
\includegraphics[width=0.8\linewidth]
{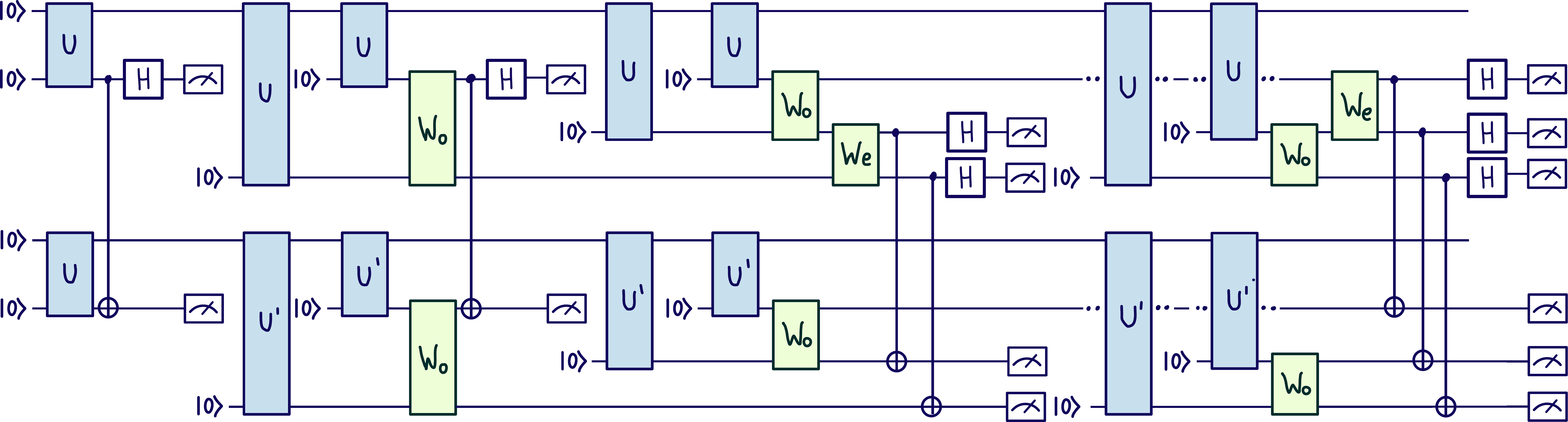}\\
\hspace*{-15cm}(b) \\ 
\includegraphics[width=0.6\linewidth]
{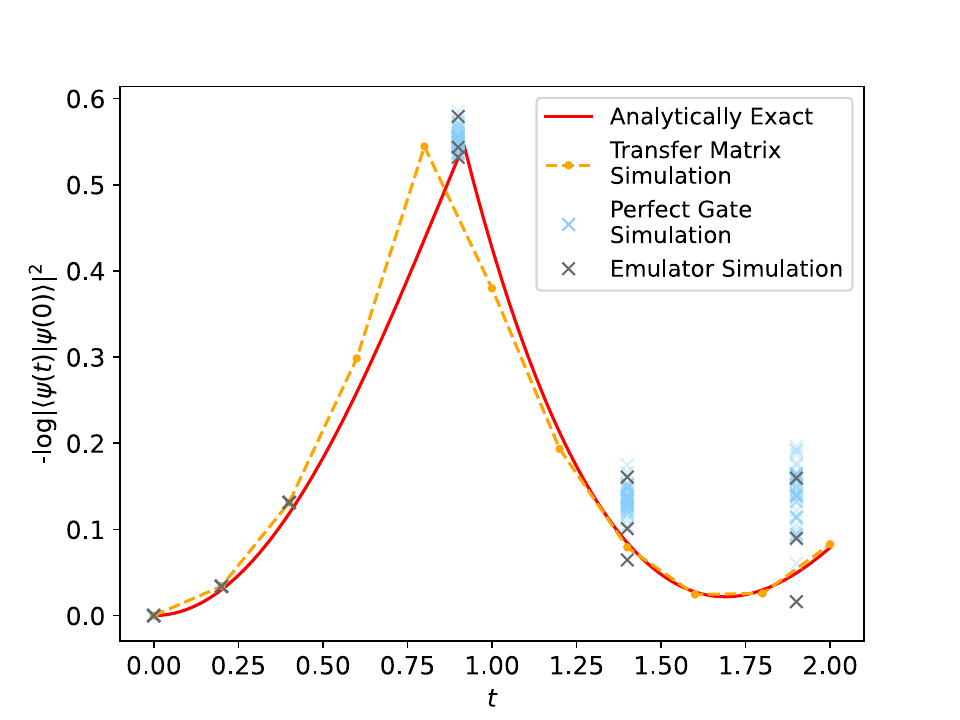}
\end{center}
    \caption{{\bf 2nd order Trotterisation:} a) The circuit calculates the cost function using a second order Trotterisation. Here the $W_o$ and $W_e$ gates are the odd and even Trotterised evolution operators.
    b) Simulation of a second-order Trotterisation scheme with $dt$=0.5. This simulation runs with a $\sim 20\%$ reduction in sampling cost. We compare the noiseless simulation (blue) with perfect gates and the emulator (grey), which includes gate noise.}

    \label{fig:HigherTrotter}
\end{figure*}

One such parameter is the order of Trotterisation chosen for the experiment. The main results from Figure \ref{fig:CentralResults} has an error equivalent to first-order Trotterisation in the time evolution operators. However, the high gate fidelities allow for experimenting with a higher order of Trotterisation. Fig. \ref{fig:HigherTrotter} shows the circuits and results of simulation for a second order Trotterisation. Due to the use of a higher order Trotterisation, the time step can be increased to $dt=0.5$. As the step size has increased, fewer steps are required to evolve the state to a given time $t$, resulting in an approximate reduction in sampling cost of $\sim 20\%$ for similar performance.

\section{Gauge and Reparametrisation Invariance}
\label{app:SlowEvolution}
A combination of gauge invariance and reparametrisation invariance can be used to reconcile the data obtained from optimisation on the quantum processor and by classical optimisation of the principal eigenvalue of the transfer matrix. 

{\it Gauge Invariance:}
The gauge invariance of iMPS states under unitary rotations in the auxilliary space,
\begin{equation}
\includegraphics[width=0.6\linewidth]
{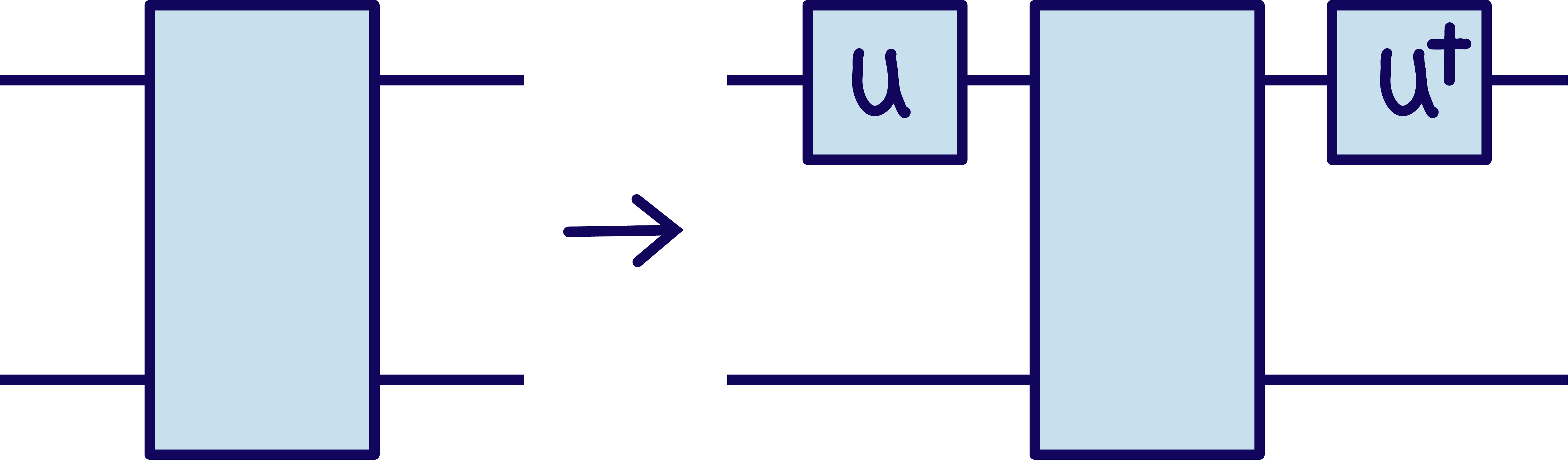}, 
\label{eq:GaugeTransI}
\end{equation}
is reduced by our restricted parametrization. Exact gauge rotations are restricted to residual $x$-gauge rotations.
An $x$-gauge rotation by an angle $\theta$ transforms ansatz parameters $\phi_0,\phi_1,\phi_4,\phi_7$ to parameters $\phi_0',\phi_1',\phi_4',\phi_7-\theta$ according to 
\begin{equation}
R_x(\theta) R_z(\phi_0') R_x(\phi_1') R_z(\phi_4') =
R_z(\phi_0')R_x(\phi_1') R_z(\phi_4').
\label{eq:GaugeTransII}
\end{equation}
Other gauge rotations can only be realised approximately with a fidelity that varies depending upon the state. 
\begin{figure*}
\hspace{-5cm}(a) \hspace{5cm}(b) \hspace{5cm}(c) \\
\includegraphics[width=0.3\linewidth]{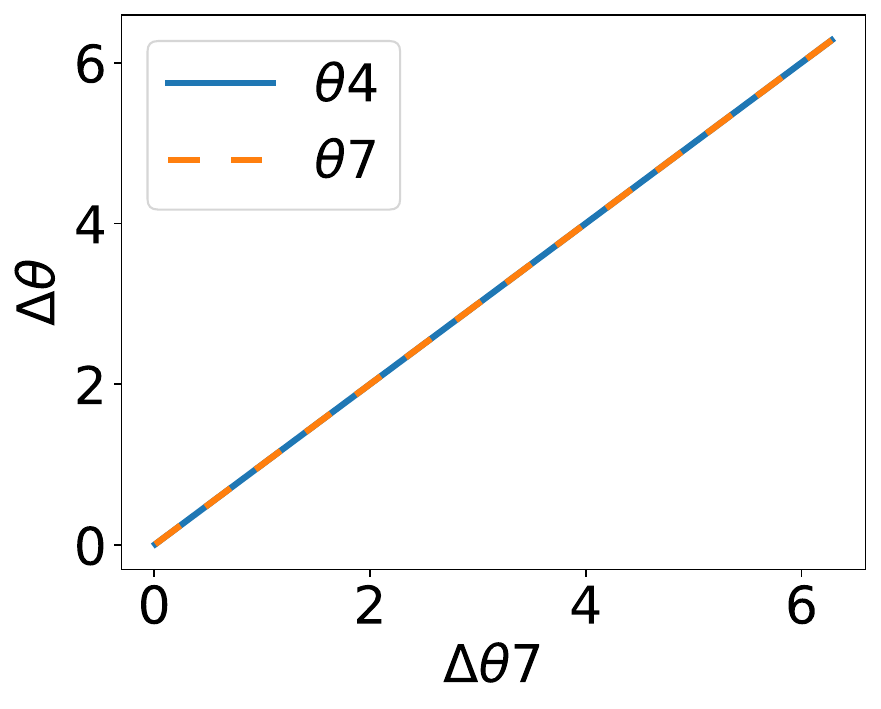} 
\includegraphics[width=0.345\linewidth]{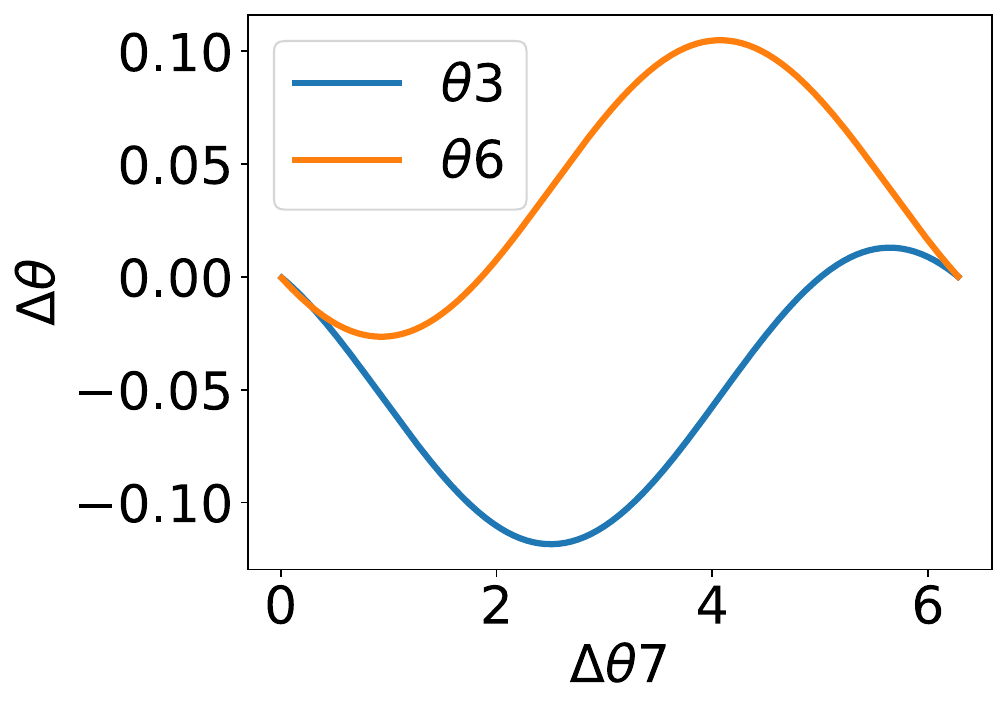} 
\includegraphics[width =0.3\linewidth,height=0.258\linewidth]
{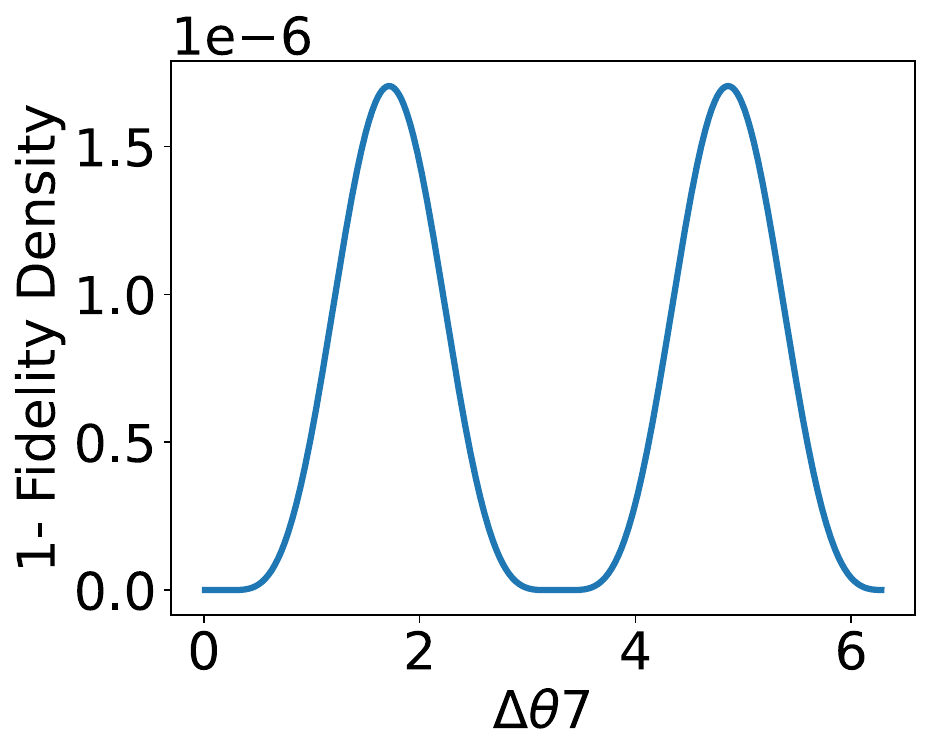}
    \caption{{\bf Parametrization invariance of the quantum  circuit MPS Ansatz:} The circuit ansatz illustrated in Fig. 1b harbours a parametrization invariance.\\a) Parameters $\theta_4$ and $\theta_7$ are varied together between $0$ and $2 \pi$. b) Parameters $\theta_3$ and $\theta_6$ vary with the change in $\theta_7$ as shown. The remaining parameters are kept fixed. c) The change in fidelity with this modification in parameters is remarkably small. }
    \label{fig:ParamInv}
\end{figure*}
\begin{figure*}
    \centering    
    \includegraphics[width=\linewidth]
    {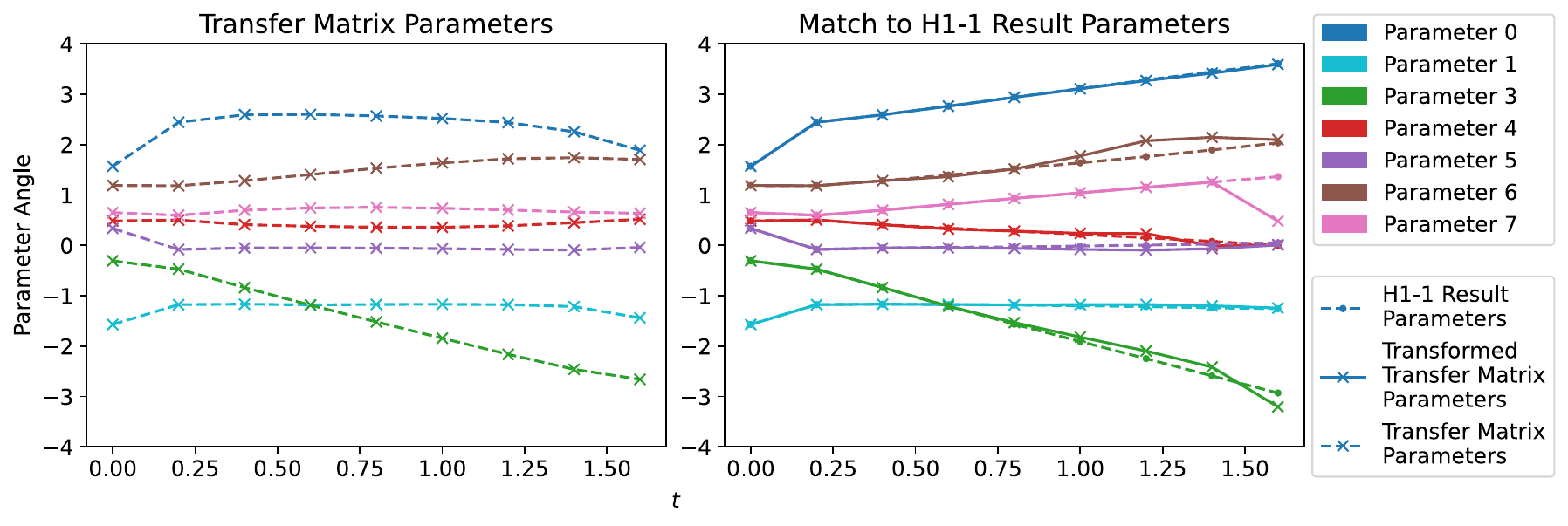}
    \caption{{\bf Mapping classically optimised data to Quantum Data:} a) Evolution of variational parameters found by classical optimisation of the mixed transfer matrix between the time-evolved MPS and the MPS with updated parameters over $10^6$ iterations of SPSA. b) Variational parameters found by optimisation on the Quantinuum H1-1 device. The classically optimised data transformed with an approximate gauge transformation and reparametrisation align well with these data.  }
\label{fig:parameter plots}
\end{figure*}

{\it Reparametrisation Invariance:} Alongside this gauge freedom, the ansatz also harbours an approximate reparametrization invariance. As illustrated in Fig. \ref{fig:ParamInv}, different sets of parameters describe the same state, in the same gauge, with high fidelity. Although still approximate, and varying with the state,   
for the states encountered in our time evolution this fidelity is very high, varying by less than $1.5\%$ and becoming as small as $\sim 10^{-6}$ at the fifth time-step. 

We obtain this approximate reparametrisation invariance in the following manner:
\\
i. We first construct the mixed transfer matrix between the parametrized unitary and its reparametrisation,
\begin{equation}
\includegraphics[width=0.3\linewidth]
{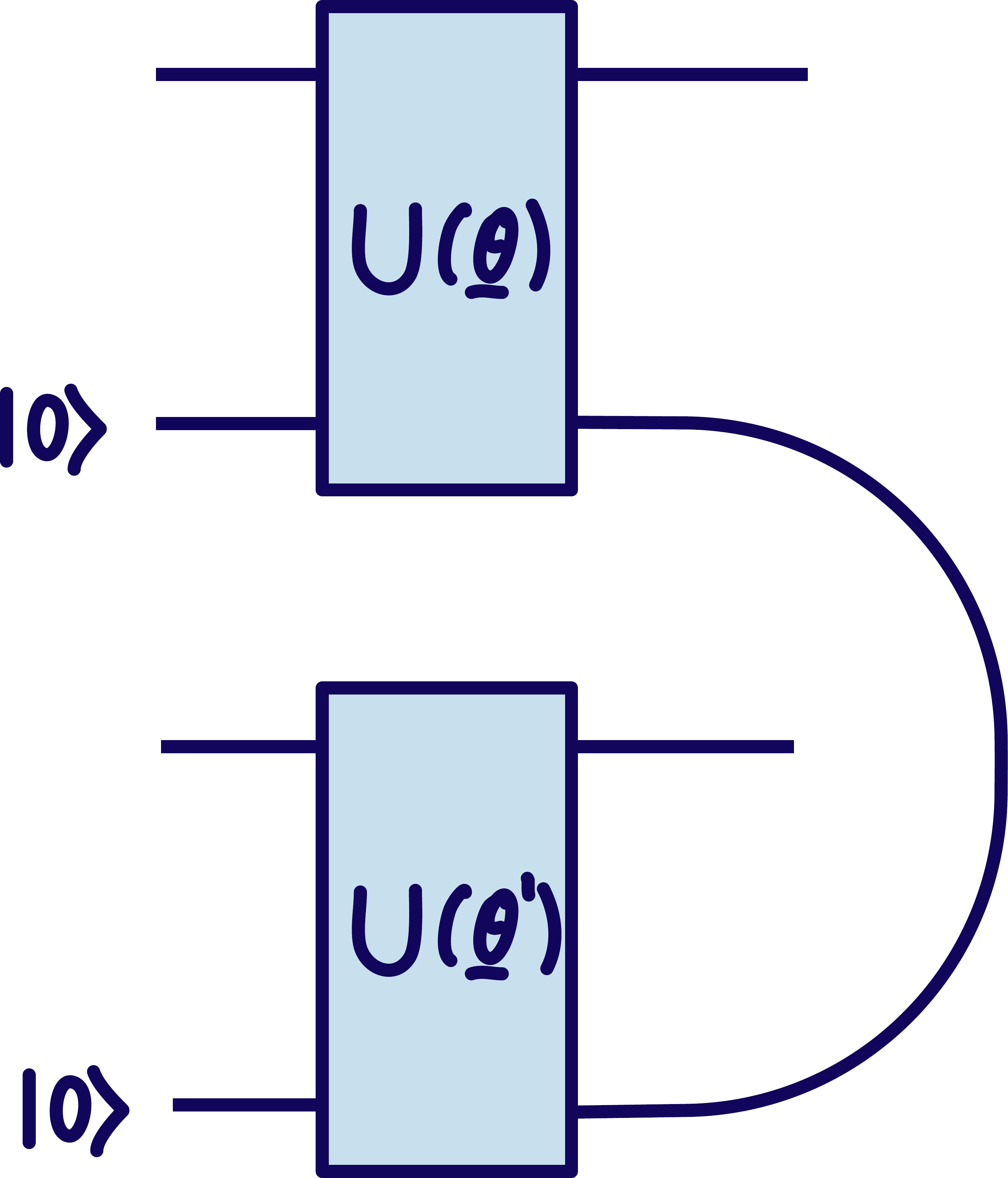}. 
\end{equation}
ii.
If the state is invariant under the reparametrization ${\bm \theta} \rightarrow {\bm \theta}'$ then the identity is preserved as a right eigenvector, {\it i.e.}
\begin{equation}
\includegraphics[width=0.6\linewidth]
{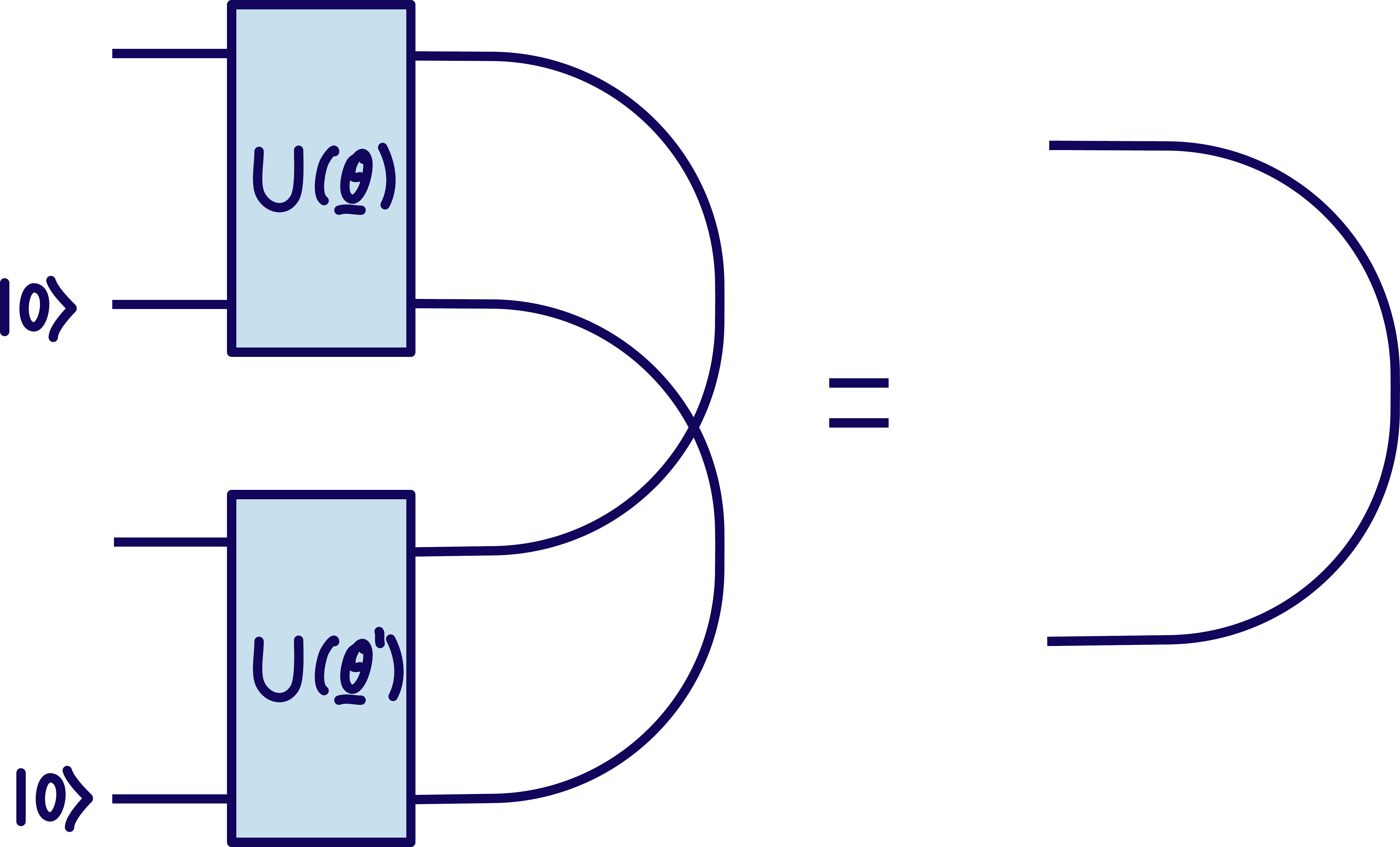}, 
\end{equation}
or equivalently
\begin{equation}
\includegraphics[width=0.7\linewidth]
{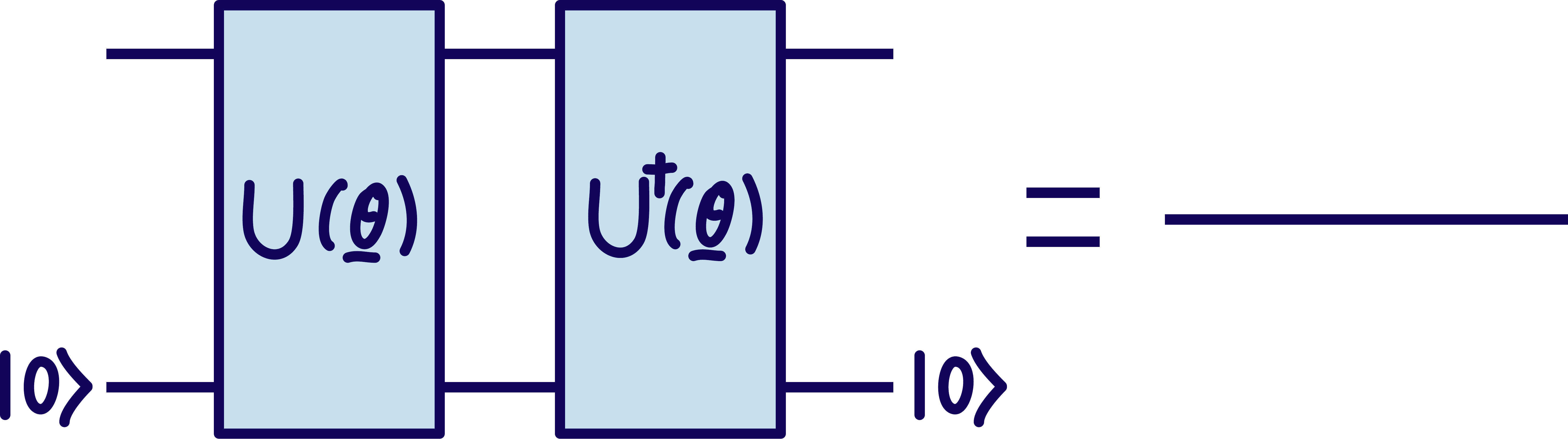}. 
\label{eq:MixedTransIdentity}
\end{equation}
\\
iii. We take the analytical expressions for the left-hand side of Eq.~(\ref{eq:MixedTransIdentity}) using our ansatz. The resulting matrix is diagonal. Setting the real parts of the diagonal elements to be precisely the identity gives a (messy) set of equations whose solutions are plotted in Fig.~\ref{fig:ParamInv}.
\\
iv.
The imaginary parts of the diagonal elements are not quite zero - indicating that, whilst it holds to a remarkable degree of accuracy, the reparametrization is only approximate. 

\vspace{10pt}
{\it Reconciling Classically Optimised and Quantum Data:}
Combining gauge rotation with reparametrisation invariance, 
there are many smooth paths in parameter space that describe the same evolution of the quantum state.
This enables us to reconcile the data from our run on the quantum computer with data obtained by full classical optimisation of the principal eigenvector of the transfer matrix (we refer to this as the {\it transfer matrix data}) and moreover to identify the main source of error in these results.
\begin{figure}[!b]
    \centering
\includegraphics[width=0.9\linewidth]
{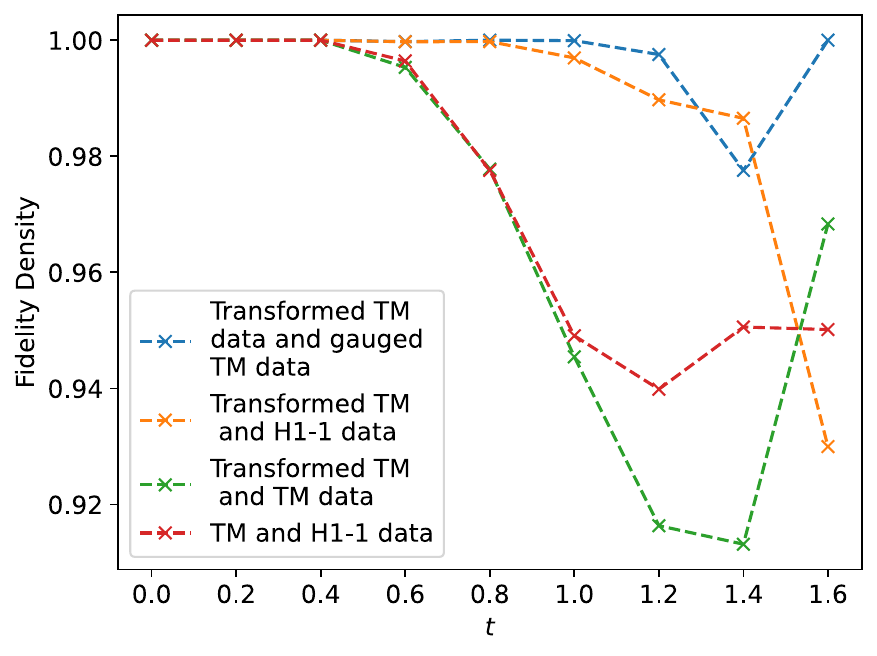}
    \caption{{\bf Fidelity between quantum and classical data and transformations between them:} The fidelity between the transfer matrix and Quantinuum H1-1 data is similar to that between the transfer matrix date and transformed (gauged and reparametrised) transfer matrix data. This suggests that most of the errors in the Quantinuum H1-1 data are explained by it settling into a gauge that is only approximately captured in our parametrisation. The fidelity between the transfer matrix data and the transformed transfer matrix data is high and similar to the fidelity between the transformed transfer matrix data and the Quantinuum H1-1 data.  }
\label{fig:fidelities}
\end{figure}
Fig.~\ref{fig:parameter plots} shows the transfer matrix data and its transformation using a gauge and reparametrisation invariance to match the quantum data. To obtain these results, we first identified the gauge transformation between the data by finding the eigenvector of the mixed transfer matrix. Following this, the residual difference was accommodated by a reparametrisation. 

Fig.~\ref{fig:fidelities} shows the fidelities after various transformations of the transfer matrix data. The bulk of the error arises because the optimised quantum data has settled into a gauge that is related to the global optimum by an approximate gauge rotation; the fidelity of the gauge rotation (seen by comparing the transfer matrix data with the gauged transfer matrix data) is about the same as that between the transfer matrix data and the quantum data. The reparametrisation error is small compared to this and essentially negligible compared to the gauge error over most of the evolution. 

The resulting linear variation of parameters 
suggests that the dynamical phase transition can be understood (to high accuracy) as an effective precession in entanglement space along the same lines as the recent MPS treatments of many-body quantum scars \cite{PhysRevX.10.011055}.

\end{document}